\documentstyle[aps,twocolumn,floats]{revtex}
\input epsf.tex

\begin{document}
\twocolumn[
\hsize\textwidth\columnwidth\hsize\csname@twocolumnfalse\endcsname

\title{Strong correlations in low dimensional conductors.\\ What are they, and
where are the challenges?}
\author{A.-M.S. Tremblay$^*$, C. Bourbonnais$^*$ and D. S\'en\'echal}
\address{D\'epartement de physique et Centre de recherche sur les
propri\'et\'es \'electroniques de mat\'eriaux avanc\'es, \\
$^*$ Institut canadien de recherches avanc\'ees,\\ Universit\'e de
Sherbrooke, Sherbrooke,\\ Qu\'ebec, J1K 2R1, Canada}
\date{April 2000}
\maketitle
\begin{abstract}%
This paper is written as a brief introduction for 
beginning graduate students. The picture of electron waves moving in a
cristalline potential and interacting weakly with each other and with
cristalline vibrations suffices to explain the properties of technologically
important materials such as semiconductors and also simple metals that
become superconductors. In magnetic materials, the relevant picture is that
of electrons that are completely localized, spin being left as the only
relevant degree of freedom. A number of recently discovered materials with
unusual properties do not fit in any of these two limiting cases. These
challenging materials are generally very anisotropic, either quasi
one-dimensional or quasi two-dimensional, and in addition their electrons
interact strongly but not enough to be completely localized. High
temperature superconductors and certain organic conductors fall in the
latter category. This paper discusses how the effect of low dimension leads
to new paradigms in the one-dimensional case (Luttinger liquids, spin-charge
separation), and indicates some of the attempts that are being undertaken to
develop, concurrently, new methodology and new concepts for the
quasi-two-dimensional case, especially relevant to high-temperature
superconductors.
\end{abstract}

\bigskip
]
\narrowtext
\tightenlines

\section{Introduction}

Quantum mechanics and statistical mechanics have provided us with the tools
to understand the behavior of bulk matter. Nevertheless, except in the case where
particles are independent, the problem of treating $10^{23}$ electrons is
unmanageable by brute force application of the basic laws. A few concepts,
and their mathematical implementation, were needed to enable us to develop
both the qualitative and highly quantitative theories that nowadays explain the
electronic and magnetic properties of solids. The computer revolution is in
part the outcome of this understanding and of the massive experimental
effort devoted to controlling semiconducting and magnetic materials, which
are the basic elements of transistors, magnetic storage materials and other
pieces of basic hardware.

What is there left to do then? All these successes may seem to indicate that
we have the tools to understand the electronic and magnetic properties of
any piece of solid matter. This is not so. This short overview, written
primarily for the student with a first course in Solid State Physics, will
summarize the traditional views of Solid State systems and move on to show
how these views fail in a large class of materials. As we will see,
high-temperature superconductors are one of the most famous examples of
materials begging for understanding. But there are others. And the mysteries
lay not only in the origins of the superconductivity itself, but also in
normal state properties that one would have expected traditional Solid State
Physics to explain. In fact, the failures of present day Solid State theory
provide an intellectual challenge of the highest level. The Physics of
strong electron-electron interactions and of systems in low-dimensional
spaces (one or two dimensions) is what is at stake. New concepts have
already emerged. For example, we know now that in one dimension,
single-electron momentum states are very bad representations of the true
eigenstates, which are, instead, collective charge and spin excitations. In
other words, in a one-dimensional solid, the electron splits into its spin
and its charge degrees of freedom. This concept of spin-charge separation is
only one example of the kind of new ideas, and corresponding tools that need
to be developed. The close interplay between experimental facts and the
developement of new qualitative ideas, as well as the occasional need for
heavy mathematical and numerical artillery, are characteristics of the field
that we also wish to illustrate.

Although there is a vast number of topics in strongly correlated electron
Physics, some of which have already led to Nobel Prizes, we will concentrate
on those topics that we have worked on and that are closely related to the
high-temperature superconductors, the subject of this issue of Physics in
Canada. We first recall the standard approaches, show experimental results
that are unexplicable within these schemes, briefly give general theoretical
arguments that tell us why the standard approaches are expected to fail in
these cases and conclude with remarks on theoretical models and
new methods that are being developed. In a nutshell,  it should be
clear at the end of this review that ``Strongly Correlated electrons'' 
refers to a rather broad class of problems originating basically from either
strong interactions or singular  scattering processes in low dimension.

\section{The standard approaches: quasiparticles and localized spins}

The groundwork for successful theories of the electronic and magnetic
properties of solids began in the early days of quantum mechanics. Bloch's
theorem explained why single-electron eigenstates in periodic arrays are
plane-wave like and can be described by a wavevector and an other quantum
number, called the band index. In the case of magnetic insulators, the
so-called Heisenberg model described the interactions of localized electrons
interacting with each other through their spin degrees of freedom. These two
dramatically opposite points of view, of delocalized vs localized electrons,
have been usefully developed and applied to different types of materials.
The present section illustrates the main concepts that have emerged.

%~~~~~~~~~~~~~~~~~~~~~~~~~~~~~~~~~~~~~~~~~~~~~~~~~~~~~~~~~~~~~~~~~~~~~~~~~~~~
\begin{figure}
\centerline{\epsfxsize 6cm \epsffile{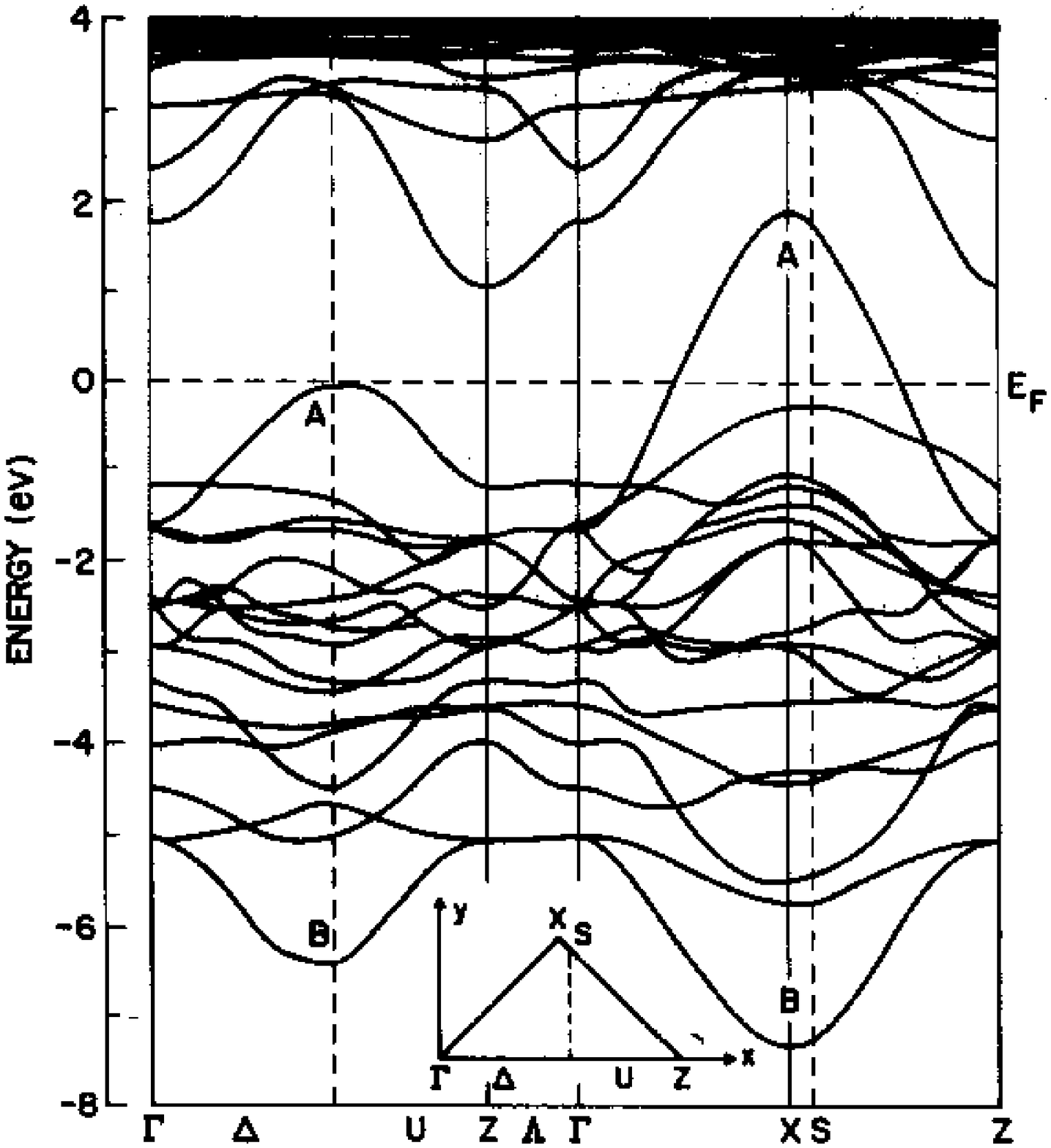}}%
\caption{Band structure of  $\rm La_2CuO_4$, taken from
Ref.~\protect\cite{Mattheiss}}%
\label{LDA}
\end{figure}
%~~~~~~~~~~~~~~~~~~~~~~~~~~~~~~~~~~~~~~~~~~~~~~~~~~~~~~~~~~~~~~~~~~~~~~~~~~~~
\subsection{Quasiparticles, the Fermi surface and Fermi liquid theory}

A mean-field (i.e. average) treatment of interactions leads to a picture
where the many-electron wavefunction is simply an antisymmetrized product
(because of Fermi statistics) of one-particle eigentates of the type
described by Bloch. The best available method to obtain these one-particle
eigenstates today is the local density approximation (LDA), a
method based on ``density functional theory''. The 1998 Nobel
Prize in Chemistry was awarded for the development of the latter
approach \cite{Kohn99}. Fig.~\ref{LDA} gives the single-particle
eigenenergies resulting from a LDA calculation for a compound in the
high-temperature superconductor family,
$\rm La_{2}CuO_{4}.$ The horizontal axis represents wavevector along
different directions and the different curves for the same wavevectors
represent the bands. The zero-temperature many-body state is built by
filling-in the lowest energy states, following the constraints of the Pauli
principle, until all electrons are accounted for. Hence, one would expect
that one can draw a surface, in wavevector space, that separates filled from
unfilled states. That is the so-called Fermi surface. When bands are either
completely filled or completely empty, one has an insulator (or a
semiconductor when the energy gap between the highest filled band and the
lowest empty band is not too large). Otherwise, there are zero-energy
excitations and the system conducts. Given that the last filled level in a
metal lies in a band that spreads over a few eV in energy, room temperature
(1/40eV) corresponds to energies that are minuscule on that scale and hence
often beyond the precision of LDA calculations. Nevertheless, these small
energies are often large enough to play an essential role in the observable
properties of the system.

Suppose we put back the interaction, i.e. we compute matrix elements of the
full Hamiltonian in the basis obtained from the LDA calculation. The
Hamiltonian in that basis contains residual interactions between
quasiparticles. Even if we cannot in practice carry out this program, the
general form of the Hamiltonian is pretty clear on the basis of general
considerations and on symmetry arguments. The residual
interactions should be short-range since the LDA bands have taken screening
into account for the most part. Furthermore, a wavefunction made of a
single antisymmetrized product of states is not an eigenstate of the
Hamiltonian that includes interactions. In a physical picture, particles
scatter off each other, changing momentum and band quantum numbers. However,
the Pauli principle strongly constrains phase space for final states. In
fact, it can be proven to all orders in perturbation theory (assuming that
it converges, which is not the case in one dimension or for strong
interactions) that even in the presence of electron-electron interactions,
the single-particle {\it excitations } near the Fermi surface are well
described by a single-particle picture. These excitations are called
quasiparticles. This is the first step in the so-called Landau Fermi liquid
theory of metals. However complicated the band structure, the arguments
given above suggest that for low-energy fermionic excitations, only the
band near the Fermi surface is relevant, giving a conceptual framework to
understand wide classes of materials. What will change from one material to
the other is the effective mass of the excitations, or more generally,
details of their energy dispersion, but the qualitative picture is quite
universal.

%~~~~~~~~~~~~~~~~~~~~~~~~~~~~~~~~~~~~~~~~~~~~~~~~~~~~~~~~~~~~~~~~~~~~~~~~~~~~
\begin{figure}
\centerline{\epsfxsize 5cm \epsffile{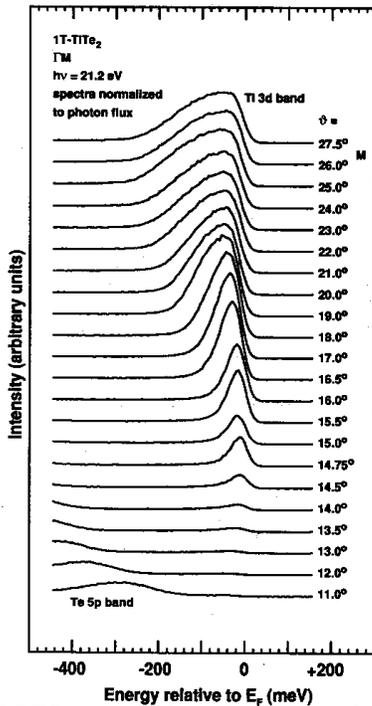}}%
\caption{ARPES spectra of $1-T-\rm TiTe_2$, taken from Fig.~1 of
Ref.~\protect\cite{Claessen92}}
\label{fig2}
\end{figure}
%~~~~~~~~~~~~~~~~~~~~~~~~~~~~~~~~~~~~~~~~~~~~~~~~~~~~~~~~~~~~~~~~~~~~~~~~~~~~
Nowadays, one can see the quasiparticles experimentally in a rather direct
manner. Indeed, synchrotron radiation has given us X-ray sources that are
powerful enough to do Angle Resolved Photoemission Spectroscopy (ARPES). In
these experiments, it is possible not only to measure the energy of the
outgoing electron, it is also possible to resolve its momentum parallel to
the surface, which is conserved when the electron is extracted by the X-ray
from the material. For a system where energy eigenstates have a strong
two-dimensional nature, that is all the quantum numbers that we need. 
Fig.~\ref{fig2} presents the results for a compound that behaves as expected
from the quasiparticle picture. The different curves correspond to different
momenta. They give on the vertical axis a quantity proportional to the
probability, times a Fermi function, that an electron of given momentum has
the energy indicated on the horizontal axis. As one moves in the various
directions of wavevector space, one reaches a point where the maximum
intensity is very near zero energy ($14.75^{0}$ on the figure). The effect
of the Fermi function is that for a probability that would be maximum {\it
at} zero energy, the observed maximum is slightly {\it below} zero energy.
At zero energy, the observed function is smaller than the value it would
have had but it still has sizeable weight. In this way, one can thus map the
wavevectors where there are no-longer electrons to photoexcite. This is the
Fermi surface (Fermi line in $d=2$). If there were no interaction between
electrons, there would be only one energy allowed for a given momentum
state. Clearly, here the probability for a given momentum is centered at a
wave-vector-dependent position but it is nonvanishing for several energies.
The width in energy for a given momentum cannot be accounted for
simply by experimental resolution. Its existence is expected from the fact
that quasiparticles have a lifetime. One can also check from the figure that
the width in energy is becoming narrower as one approaches the Fermi
surface. Also, although this is not shown here, at the Fermi surface the
width decreases rapidly with temperature (like $T^{2}$).

\subsection{Thermodynamics of the Fermi liquid, and phase transitions}

So much for single-particle properties. What happens when one measures
thermodynamic quantities such as the magnetic susceptibility or the specific
heat? One is not adding or removing new particles in the system. The external
probe is just emptying occupied states while filling unoccupied states: It
is creating particle-hole excitations. In that case, Landau Fermi liquid
theory predicts that the interactions have a mean-field-like effect that
modifies the predictions one would obtain for noninteracting particles. For
example, the specific heat is linear in temperature and proportional to the
density of single-particle excitations, like in a free electron gas. The
spin susceptibility should then be temperature independent, like the Pauli
susceptibility of free electrons, and, in a naive picure, it should also be
proportional to the density of single-particle excitations. In reality, Fermi
liquid theory  tells us that there is an enhancement factor $\left(
1+F_{0}^{a}\right) ^{-1},$ where
$F_{0}^{a}$ is a measure of the interaction$.$ 

Interactions, quite generally, are the cause of phase transitions. The case
$F_{0}^{a}=-1$ above corresponds to $\left( 1+F_{0}^{a}\right)
^{-1}=\infty,$  which means a divergent static spin susceptibility. That is
a clear signal for the onset of ferromagnetism. The ferromagnetic state
breaks spin rotational invariance. Another example of phase transition
caused by interactions, is the superconducting transition, which breaks
global gauge invariance. This is discussed further by J. Carbotte in this
issue. The origin of the interactions leading to phase transitions may be
quite subtle. For example, in the case of conventional superconductors, the
retarded electron-phonon interaction leads to an effective attraction
between quasiparticles that is ultimately responsible for the
superconducting state. Finding and characterizing all possible states of
matter caused by interactions is a field of endeavour in itself.

\subsection{What about the Heisenberg model?}

While band theory works well with most  materials
with only $s$ and $p$ derived bands,  in certain cases, mostly with $d$ and
$f$ electron materials, it is totally inappropriate. One of the most famous
examples is $\rm V_{2}O_{3}.$ The band structure predicts it should be a
metal. Instead, at low pressure and low temperature it is an
antiferromagnetic insulator. In other words, electrons do not move (that
defines the insulator) unless they are kicked really hard, and spins, on the
other hand, order in an up-down pattern on alternating sites (that is called
an antiferromagnet). As we shall discuss more later, this failure of band
theory comes from the fact that interactions in these materials are larger
than kinetic energy effects, leading to a breakdown of perturbation theory.
In the jargon, this is a ``strong coupling'' effect. Despite the
difficulties of deriving from first principles the Hamiltonian in the
reduced Hilbert space involving only spin degrees of freedom of the last
filled states, again one can use general symmetry principles to write down
model Hamiltonians. Solving these is a non-trivial task that has been
successfully accomplished by many people for many years in the field of
magnetism. Numerous neutron scattering experiments have verified in details
the predictions of these models in many cases.

\section{Experimental evidence for failures of standard Solid State
theory in low-dimensional conductors.}

What do we mean by low dimensional conductors? In practice they can be
formed, for example, by organic molecules stacked onto each other, or by
copper oxygen planes separated by ions, as in the case of high-temperature
superconductors. Despite the horrendous complexity of these structures, the
Pauli principle and the general arguments given above tell us that for
low-energy Physics we can concentrate only on the LDA bands that are very
close to the Fermi energy. It turns out that, in many realizable
cases, there is only one such band. Furthermore, the eigenstates in that
band may turn out to be very different depending on which
axis one is looking from. In these very anisotropic cases, it is as if
electrons moved preferentially in one or two dimensions, the latter being
the case for the high-temperature superconductors. Let us see what non-Fermi
liquid Physics can arise in the $d=1$ and $d=2$ cases.

%~~~~~~~~~~~~~~~~~~~~~~~~~~~~~~~~~~~~~~~~~~~~~~~~~~~~~~~~~~~~~~~~~~~~~~~~~~~~
\begin{figure}
\centerline{\epsfxsize 7cm \epsffile{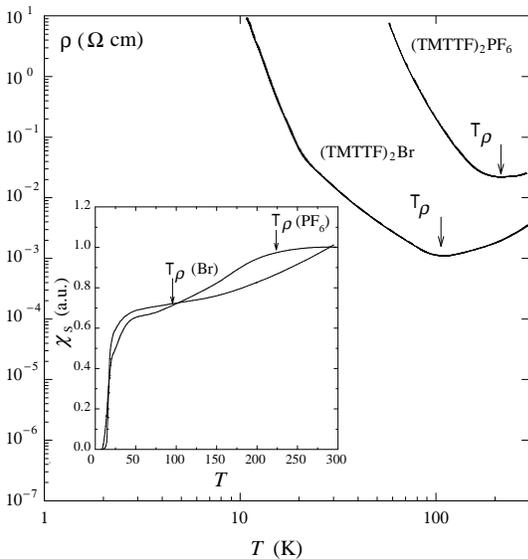}}%
\caption{Electrical resistivity as function of temperature for two members
of the (TMTTF)$_2$X series. Inset: Temperature-dependent spin
susceptibility, from Ref.~\protect\cite{RevueCBjerome99}.}
\label{rho-chi}
\end{figure}
%~~~~~~~~~~~~~~~~~~~~~~~~~~~~~~~~~~~~~~~~~~~~~~~~~~~~~~~~~~~~~~~~~~~~~~~~~~~~

\subsection{One dimension: spin-charge separation in the organics}

When electrons of opposite momentum are confined to move in  one
spatial direction, they cannot avoid each other and  their interaction will
be in  some way enhanced in comparison with  isotropic systems. As we will
explain in the theory section, quasiparticles are absent in one dimension,
and one has instead a Luttinger liquid where  harmonic collective
oscillations of both spin and charge are the true elementary excitations.

Here we present two clear experimental examples of the failure of the
quasiparticle picture.  Consider the normal phase of the (TMTTF)$_2$X series
of quasi-one-dimensional organic conductors  (here, TMTTF stands for the
tetramethylfulvalene molecule and X= PF$_6$, Br, ..., for an inorganic
monovalent anion) \cite{RevueCBjerome99}. As shown in Fig.~\ref{rho-chi},
there ia a clear upturn in electrical resistivity at temperature $T_\rho$,
which depicts a change from metallic to insulating behavior. Below
$T_\rho$, charge carriers become thermally activated. In a band picture of
insulators, the same thermally activated behavior should be present for
spins since the only way to create spins in a band insulator is to excite
quasiparticles across the gap between the filled and the empty bands. For
the compounds shown in Fig.~\ref{rho-chi}, spin excitations instead are
unaffected and remain gapless. This is  shown by the regular temperature
dependence of the spin susceptibility $\chi_s$ at $T_\rho$ (inset of
Fig.~\ref{rho-chi}).

%~~~~~~~~~~~~~~~~~~~~~~~~~~~~~~~~~~~~~~~~~~~~~~~~~~~~~~~~~~~~~~~~~~~~~~~~~~~~
\begin{figure}
\centerline{\epsfxsize 8cm \epsffile{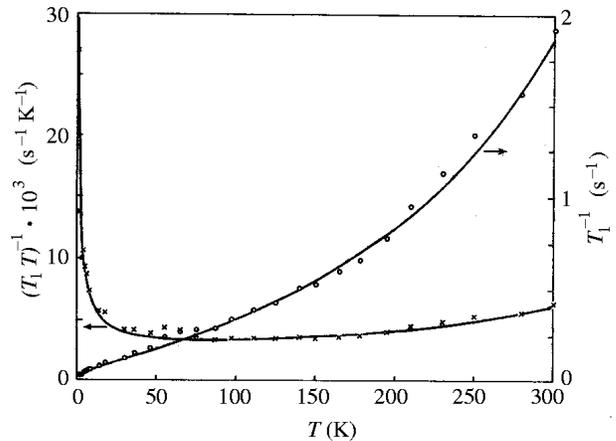}}%
\caption{Temperature dependence of $(T_1T)^{-1}$ ($\times$) and $T_1^{-1}$
($\circ$) for TTF[Ni(dmit)$_2$]$_2$. The continuous line corresponds to the
Luttinger liquid prediction, from Ref.~\protect\cite{Bourbon88}.}
\label{dmit}
\end{figure}
%~~~~~~~~~~~~~~~~~~~~~~~~~~~~~~~~~~~~~~~~~~~~~~~~~~~~~~~~~~~~~~~~~~~~~~~~~~~~

Among other experimental tools that are quite useful in probing
signs of unusual behavior in low-dimensional organic conductors is Nuclear
Magnetic Resonance, especially the temperature dependence of the nuclear
spin-lattice relaxation rate, denoted as
$T_1^{-1}$. Nuclear and electronic spins being   coupled through the
hyperfine interaction, the measurement of $T_1^{-1}$ can give valuable
information about electronic spin excitations. While $ (T_1T)^{-1}$ is
temperature-independent in a Fermi liquid, the correct theory in one
dimension for
$(T_1T)^{-1}$  takes the form \cite{Bourbon93}:
\begin{equation} (T_1T)^{-1} = C_1T^{K_\rho-1} + C_0 \chi_s^2(T),
\label{relaxation}
\end{equation} where the exponent $K_\rho \ge 0$ stands for the `stiffness'
constant of collective charge degrees of freedom.    It gives rise to a
power-law enhancement of $ (T_1T)^{-1}$, which comes from antiferromagnetic
spin  correlations. For one-dimensional  insulating compounds like
(TMTTF)$_2$X, charge degrees of freedom are frozen so that 
\hbox{$K_\rho=0$}.  The resulting behavior $ T_1^{-1}\sim C_1 + C_0\chi_s^2$
turns out to be  invariably found in all these insulating materials down to 
low temperature, where three-dimensional magnetic or lattice long-range
order is stabilized \cite{Bourbon93}.  Among the very few
quasi-one-dimensional organic materials that do not  show  long-range
ordering, the  case of TTF[Ni(dmit)$_2$]$_2$ is
interesting \cite{Bourbon88}: This system  remains metallic down to very low
temperature and a power law enhancement ($K_\rho \approx  0.3$) of
$ (T_1T)^{-1}$ is maintained from $300 K$ down to $1 K$ or so
(Figure~\ref{dmit}).

\subsection{Two dimensions: The pseudogap.}

\label{SecVanishingAct}

We have already shown the band structure of $\rm La_{2}CuO_{4}$ in
Fig.~\ref{LDA}. The last occupied band is essentially a linear combination
of copper and oxygen orbitals corresponding to two-dimensional (planar)
arrangements of $\rm CuO_{2}$ atoms. Thus, one expects that
electrons relevant for transport are essentially confined to two dimensions.
This is confirmed by the highly anisotropic transport properties of these
materials, as discussed by T. Timusk in this issue. The Fermi level crosses
the last occupied band, so we expect a metal.

But in reality, $\rm La_{2}CuO_{4}$ is an antiferromagnetic insulator! This
is because of strong interactions. When $\rm La^{3+}$
cations located away for the conducting planes are replaced by $\rm Sr^{2+}$
cations, electrons are removed from the $\rm CuO_{2}$ planes and
${\rm La}_{2-x}{\rm Sr}_{x}{\rm CuO}_{4}$ becomes eventually a high-temperature
superconductor. The generic phase diagram for high-temperature
superconductors appears elsewhere in this issue. There are many
high-temperature superconductors, and they all have in common $\rm CuO_{2}$
planes that can be doped. The physical properties of these planes are quite
similar from one compound to the next. From being antiferromagnetic when
there is one electron per $\rm CuO_{2}$ unit, they become superconductors
when doped with holes (or with electrons in certain compounds). With hole
doping, the superconducting $T_{c}$ first increases. That is called the
underdoped region. Then, a maximum $T_{c}$ is reached at ``optimal
doping'', decreasing thereafter in the ``overdoped'' region.

%~~~~~~~~~~~~~~~~~~~~~~~~~~~~~~~~~~~~~~~~~~~~~~~~~~~~~~~~~~~~~~~~~~~~~~~~~~~~
\begin{figure}
\centerline{\epsfxsize 6.5cm \epsffile{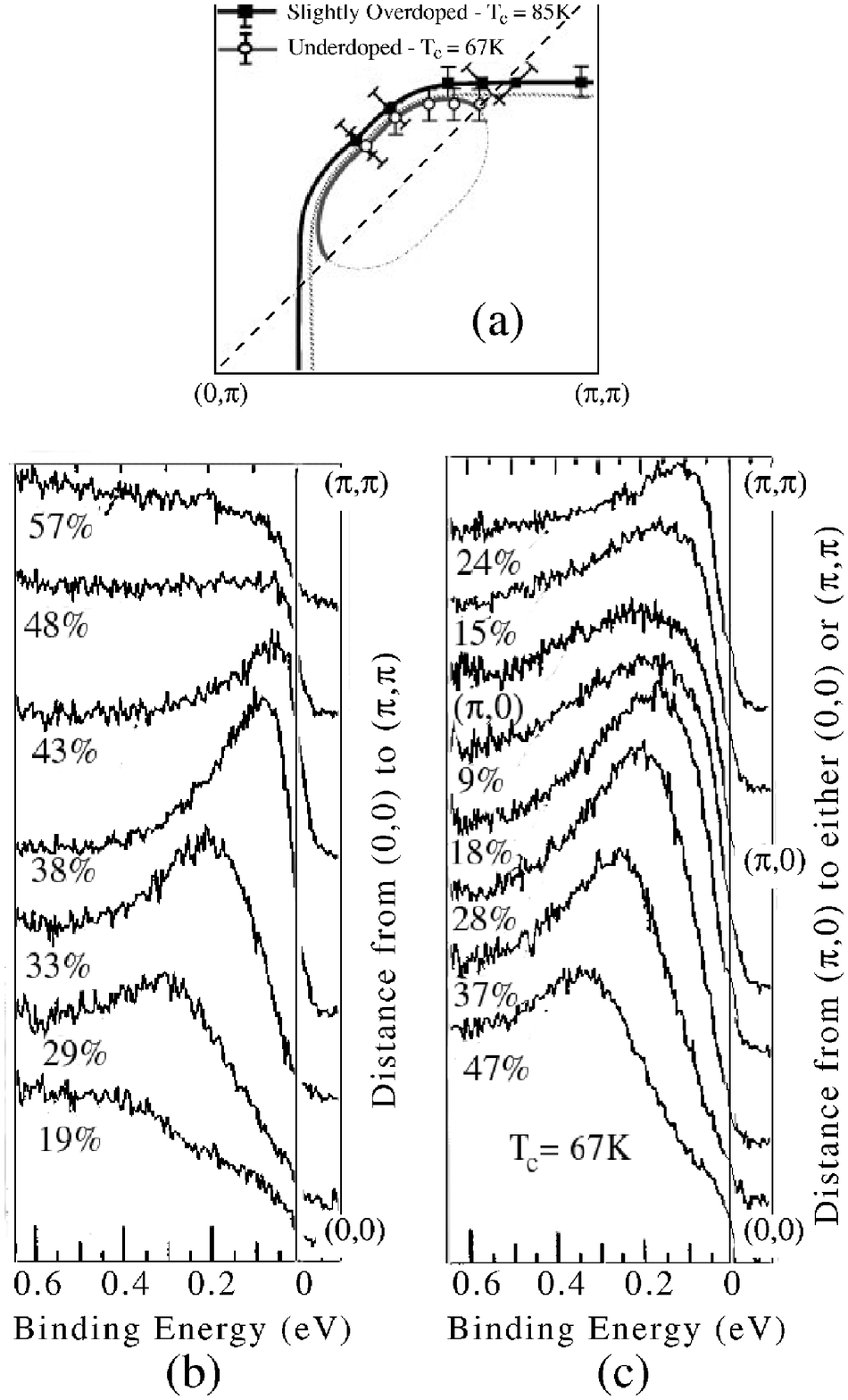}}%
\caption{ARPES spectra of $\rm O_2$-reduced $\rm
Bi_2Sr_2CaCu_2O_{8+\delta}$, taken from
Ref.~\protect\cite{Marshall96}}%
\label{fig3}
\end{figure}
%~~~~~~~~~~~~~~~~~~~~~~~~~~~~~~~~~~~~~~~~~~~~~~~~~~~~~~~~~~~~~~~~~~~~~~~~~~~~
Let us look at the underdoped regime, above $T_{c}.$ To see if the standard
Fermi-liquid approach applies in this regime, we resort to ARPES. It is
experimentally difficult to do ARPES in $\rm La_{2-x}Sr_{x}CuO_{4}$, so we
use results obtained from the $\rm CuO_2$ planes of the so-called Bi2212
high-temperature superconductor. In Fig.~\ref{fig3}(a), the solid line
shows the location of the Fermi line expected from band structure
calculations. Fig.~\ref{fig3}(b), illustrates the ARPES spectrum obtained
for various wavevectors along the $\left( 0,0\right) $ to $\left( \pi ,\pi
\right) $ direction. At the wave-vector location expected from band
structure, one finds the properties expected for a state at the Fermi
surface, namely at zero energy the photoemission intensity is a sizeable
portion of the value at the peak position. The surprise arises when one
looks along the $\left( \pi ,0\right) $ to $\left( \pi ,\pi \right)
$ direction, Fig~\ref{fig3}(c). None of the photoemission curves has the
features expected from a state at the Fermi surface. It is as if the Fermi
line had disappeared. This is the co-called pseudogap phenomenon. It is as
if an energy gap had opened on part of what should have been the Fermi line
(hence the ``pseudo'' prefix, since zero-energy excitations are left
elsewhere in wave-vector space). If you think about it from a quasiparticle picture,
this is completely crazy. Take an energy band in a two-dimensional system.
The allowed wavevectors cover a finite region of the two-dimensional
$k_{x},k_{y}$ plane. That is the Brillouin zone. Plot the energy
corresponding to a given wavevector in the $z$ direction. That gives a
singly connected surface. Now, cut this surface by a plane parallel to the
$k_{x},k_{y}$ plane. The intersection of that plane with the energy surface
can only be of two types. Either it is a line that links one edge of the
Brillouin zone to another (or the same) edge, or it is a closed line inside
the zone. The two possibilities can exist at the same time: In
other words, there may be several Fermi lines in the Brillouin zone. But
according to these simple geometrical considerations, there is no other
possibility. If you cover the Brillouin zone with ARPES measurements, you
find that in the underdoped high-temperature superconductors the Fermi line
does something worse than disagreeing with band structure calculations. It
disappears in thin air! That is a total no-no in both standard approaches.
Remember that in the standard approaches, either you have quasiparticles and
there is a Fermi line, or you have an insulator and there is no
single-particle state at all at zero energy.

Other manifestations of the pseudogap, expecially in transport, are
discussed elsewhere in this issue.

\section{Why do the standard approaches fail?}

The failure of the standard approaches in low dimension is not a total
surprise from a theoretical standpoint. On the contrary, for a long time
there have been papers discussing the peculiarities of low-dimensional
systems. For example, consider the Mermin-Wagner theorem, which
states that a spontaneous breaking of a continuous symmetry (e.g. a
rotation) cannot occur in low dimension.

To be more specific on what that means, let us give an example. In three
dimensions, Heisenberg antiferromagnets exist at finite temperature. In two
dimensions, thermal fluctuations forbid such order from occuring at finite
temperature. A rough argument for that is as follows. At long wavelengths,
the energy associated with a change in the relative angle between
neighboring spins, $\theta ,$ will be proportional to $\left( \nabla \theta
\right) ^{2},$ or $q^{2}\theta _{{\bf q}}\theta _{-{\bf q}}$ in Fourier
space. The mean square of the local angle is given by the integral over all
wavevectors of $\left\langle \theta _{{\bf q}}\theta _{-{\bf q}%
}\right\rangle .$ Using the classical fluctuation-dissipation theorem, this
means that $\left\langle \theta ^{2}\right\rangle \propto \int d^{d}q\;
(k_{B}T/ q^2)$. That integral diverges logarithmically in two dimensions,
which proves {\it ad absurdum} that long-range order cannot exist. At zero
temperature, the above argument fails and antiferromagnetic long-range order
may exist. In one-dimension, quantum fluctuations have a similar detrimental
effect and, even at zero temperature, antiferromagnetic or superconducting
long-range order does not exist.

All this classical and quantum fluctuation business is bad news for the
quasiparticle approach. Indeed, even though long-range order does not set
in, below a temperature of the order of what would have been the mean-field
transition temperature, there are collective modes that spread over large
distances, making the material appear ordered over large scales. This
strongly scatters quasiparticles, and in some instances it may lead to short
lifetimes, or even to pseudogap phenomena
\cite{Vilk97}.

So much for doing all at once. Let us consider first the
effects of low-dimension in weak to intermediate coupling, and after that
the effects of strong interactions. In weak to intermediate coupling, things
behave very differently according to dimension. Dimension still plays a role
in strong coupling, but some strong-coupling effects depend little on spatial
dimension.

\subsection{The effects of low dimension in weak to intermediate coupling}

\subsubsection{One dimension}

General considerations on phase space and the Pauli principle tell us that
in high dimension, the scattering rate of quasiparticles at the Fermi
surface is proportional to $\left( T/E_{F}\right) ^{2}.$ Since the relative
width of the Fermi function is of order $T/E_{F},$ it makes sense to expect
that thermodynamic properties will not much be influenced by the $\left(
T/E_{F}\right) ^{2}\ll T/E_{F}$ width of the quasiparticles. In one
dimension, this argument fails. The width in perturbation theory is
proportional to $T$, like the Fermi function. Right from the start, this
invalidates the Fermi-liquid starting point. In addition, response functions
diverge as $\ln T$ at low temperature.  More specifically, what makes one
dimension so special lies in the shape of the Fermi surface, which consists
of two points ($\pm k_F$). Electron and hole states that are created by
electron-electron scattering close to $\pm k_F$ lead to elementary
superconducting (Cooper) and density-wave ($2k_F$ electron-hole) pairings;
these are not only singularly enhanced at low temperature, but their
confinement in $k-$space produces  strong interferences between them that
persist to all orders in perturbation theory. A striking outcome of this
interference is an instability of the Fermi liquid towards the formation of
a quite different quantum state called a Luttinger liquid. 

The point of view has to change completely. The appropriate theoretical
tools here bear the name of renormalization group\cite {BourbonnaisCaron}
or bosonization \cite{boso}. They lead to the same final picture: It is best
to consider spin and charge collective modes as the elementary excitations.
In the resulting ``Luttinger liquid'' picture \cite{Haldane81}, which replaces
the Fermi liquid as a general limiting case in one dimension, the spin and the
charge of would-be quasiparticles separate,  becoming the true elementary
excitations that propagate at different velocities. We have illustrated experimental
manifestations of this phenomenon in organic conductors in the previous section. The
cases where the compounds were insulators displayed extreme examples of spin-charge
separation. These compounds have a commensurate band filling
and their insulating behavior is a manifestation of one-dimensional Mott
localisation, a more general topic on which we return in the discussion on
the effects of strong interactions.

\subsubsection{Two dimensions}

Contrary to the one-dimensional case, the quasiparticle picture does not
fail automatically in two dimensions. For example, the compound in Figure 2
is a two-dimensional Fermi liquid. Theoretically, in two dimensions there
are only weak logarithmic corrections to the standard phase space arguments
of Fermi liquid theory. Stronger corrections occur when the Fermi surface has
so-called nesting properties \cite{Lemay}, or when one enters a fluctuation
regime. Let us consider the latter case. The fluctuation regime may occur
over a broad temperature range in two dimensions, basically from a
temperature of the order of the mean-field transition temperature, all the
way to zero temperature. Let $\xi $ be the length over which the collective
mode fluctuations are correlated. In the fluctuation regime, the scattering
rate for quasiparticles at the Fermi surface is proportional to $\frac{T}{%
v_{F}}\int d^{d-1}q\left( q^{2}+\xi ^{-2}\right) ^{-1}\propto T\xi
^{3-d}/v_{F}.$ In $d=2,$ this becomes $\xi/\xi_{th}$, where 
$\xi_{th}\equiv\hbar v_{F}/k_B T$ is the thermal de Broglie wavelength.
Since the correlation length $\xi $ diverges much faster as $T\rightarrow 0$
than $\xi_{th}$, this
implies a divergent scattering rate. It is difficult to have a stronger
contradiction of the quasiparticle picture. Physically, when the correlation
length $\xi$ becomes much larger than the thermal de Broglie wavelength, the
quasiparticles are moving in a locally ordered background. Then a pseudogap,
precursor of the $T=0$ ordered state, opens up at the Fermi surface
\cite{Vilk97}. As temperature decreases, it may open on certain segments of the
Fermi surface before it opens on other segments. That is a consequence of the
fact that the scattering rate, proportional to
$T\xi /v_{F},$ may be very different on different parts of the Fermi
surface. Close to half-filling in particular, the Fermi velocity nearly
vanishes at certain points of the Fermi surface while it is large at other
points.

\subsection{The effects of very strong interactions}

When interactions are very strong, electrons avoid getting close to each
other by localizing. When an odd number of electrons is localized on each
atom, the charge does not move and the only degree of freedom left at low
energy is essentially the spin. Note the contrast with the quasiparticle
picture where a half-filled band is a metal. The low energy Physics in these
system, where electrons are localized by interactions, is then essentially
governed by variations of the Heisenberg Hamiltonian described above. Many
years ago, Mott imagined what would happen to a system as the strength of the
interaction is increased. The transition from extended quasiparticle states
to localized states produced by large interaction effects  is refered to as
the Mott transition. It is a first order transition whose Physics has become
better understood in recent years \cite{Dispute}, thanks to the developement
of calculational methods in the limit of infinite dimension \cite{DMF}. In
the Mott insulator, many properties are not strongly dependent on dimension,
in particular when they concern high energy. That is why
infinite-dimensional methods have been useful. Nevertheless, precursor
effects caused by collective mode fluctuations have been seen in models of
Mott insulators in low dimensions \cite{Pairault}. These effects do not occur
in infinite dimension.

%~~~~~~~~~~~~~~~~~~~~~~~~~~~~~~~~~~~~~~~~~~~~~~~~~~~~~~~~~~~~~~~~~~~~~~~~~~~~
\begin{figure}
\centerline{\epsfxsize 6cm \epsffile{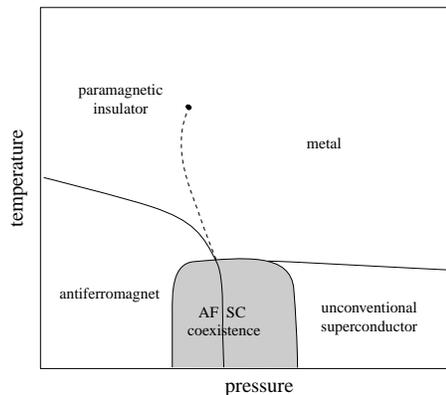}}%
\caption{Schematic phase diagram of the quasi-2D organic compound
$\kappa$-(ET)$_{2}$Cu[N(CN)$_{2}$]Cl, from
Ref.~\protect\cite{Lefebvre00}}%
\label{BEDT}
\end{figure}
%~~~~~~~~~~~~~~~~~~~~~~~~~~~~~~~~~~~~~~~~~~~~~~~~~~~~~~~~~~~~~~~~~~~~~~~~~~~~
The Mott transition does not break any symmetry, and it may occur in any
dimension. For example, $\rm  V_2O_3$ may exhibit such a transition,
although questions regarding the effects of lattice symmetry change and of
orbital degeneracies are still open \cite{Long}. A clearer example of a Mott
transition has been discovered recently in two-dimensional
organic conductors \cite{Lefebvre00}. The phase diagram is illustrated on
Fig.~\ref{BEDT}. The system is half-filled. The horizontal axis represents
pressure. From a model point of view, increased pressure means larger
overlap between atomic orbitals and hence increased kinetic energy. Indeed,
at low pressure on this diagram, the system is either a paramagnetic
insulator at high temperature, or an antiferromagnetic insulator at low
temperature. At higher pressure, one crosses a first order transition that
leads to a metallic state at high temperature and to a
$d$-wave superconductor at low temperature.

\subsection{And high-temperature superconductors in all that?}

The high-temperature superconductors are Mott insulators at half-filling.
Doping eventually leads to a $d$-wave superconducting state. Their electronic
properties are also highly two-dimensional, in particular in the underdoped
region. They thus manifest all the complexities described above. The high
energy (100 meV) pseudogap described in Section \ref{SecVanishingAct}
above, is likely to be a strong-coupling pseudogap, in other words a
pseudogap originating from the Physics of doped Mott insulators. However,
closer to the superconducting phase transition, in the more metallic regime,
one expects a fluctuation-induced pseudogap. Indeed, in photoemission, one
can often identify a lower energy pseudogap that seems to occur in a
fluctuation regime. A more detailed discussion appears in
Ref.~\cite{Moukouri00}.

\section{Theoretical methods and challenges}

One of the most widely studied model Hamiltonians of correlated electrons is the
so-called one-band Hubbard Hamiltonian:
\begin{equation} H=-\sum_{<ij>\sigma }t_{i,j}\left( c_{i\sigma }^{\dagger
}c_{j\sigma }+c_{j\sigma }^{\dagger }c_{i\sigma }\right)
+U\sum_{i}n_{i\uparrow }n_{i\downarrow \,\,\,\,}.  \label{Hubbard}
\end{equation} In this expression, the operator $c_{i\sigma }$ destroys an
electron of spin
$\sigma $ at site $i$. Its adjoint $c_{i\sigma }^{\dagger }$ creates an
electron and the number operator is defined by $n_{i\sigma }=$ $c_{i\sigma
}^{\dagger }c_{i\sigma }$. The symmetric hopping matrix $t_{i,j}$ determines
the band structure, which here can be arbitrary. Occupation of a site by
both a spin up and a spin down electron costs an energy $U$ due to the
screened Coulomb interaction. This Hamiltonian is clearly a caricature of
reality, but what is important is that it has a minimal number of parameters
and it allows one to describe the two limiting cases of delocalized and
localized electrons, as well as the Mott transition between these two
limits. Consider the case where the band is characterized by a single
parameter $t$ representing hopping between neighboring sites. At weak
coupling, when $U/t\ll 1,$ one can apply the standard quasiparticle
approach. At strong coupling, when $U/t\gg 1,$ one can show how this
Hamiltonian becomes, at low energy and half-filling, the Heisenberg
Hamiltonian for spins. Hence, it is a good starting point in both the strong and
weak coupling limits as well as in the intermediate coupling regime,
characteristic of high-temperature superconductors, where neither of the two
standard approaches work. At half-filling, the high-temperature superconductors
become antiferromagnetic insulators that are well described by the Heisenberg
model. At low energy and away from half-filling, the Hubbard model becomes a
variant of the so-called $t-J$ model, widely studied also in the context of
high-temperature superconductors.

It is hard to know from first principles if $U/t$ will be large or small for
a given system. But there are heuristic guides coming from Chemistry and
from so-called ``constrained LDA calculations''. In general, the Hubbard
Hamiltonian is an effective Hamiltonian. It is even useful in some cases to
let $U<0$ to study models of $s$-wave superconductivity.  Despite the fact
that the Hubbard model was proposed almost 40 years ago, it is only recently
that it has become to be understood at intermediate coupling and in low
dimension. Various methods have been developped to study this model. In one
dimension, an exact solution was found by Bethe Ansatz \cite{Ha96}, from
which physical information is unfortunately quite difficult to extract. The
linear dispersion of the one-dimensional electron gas in one dimension is at
the root of an analogy with relativistic field theories which explains the
success of field theoretic methods like the renormalization
group \cite{BourbonnaisCaron}, bosonization \cite{CFT} and Conformal Field
Theory\cite{CFT}. In two and more dimensions, let us mention Slave-boson
approaches \cite{SlaveBoson}, renormalized perturbation theory
approaches \cite{bickers1}, strong-coupling perturbation expansions 
\cite{Pairault} and the two-particle self-consistent approach \cite{Vilk97}.
Finally, infinite-dimensional methods have provided a dynamical mean-field
theory methodology \cite{DMF} that has been very useful in understanding
the Mott transition. This approach can also be extended to lower
dimensions. In $d=2$ however, the effect of antiferromagnetic fluctuations are
not included yet in this  methodology, which limits somewhat the
applicability of the method to high-temperature superconductors.

A major factor for progress is that it is now possible to do reliable
numerical calculations that allow us to both develop physical intuition and
check the validity of approximation methods. Exact diagonalizations are
possible in any dimension but are restricted to a small number of
electrons \cite{Senechal00}. In one dimension, Density Matrix
Renormalization Group \cite{DMRG} has provided a revolutionary method to
obtain reliable results. In two dimensions, Quantum Monte Carlo
simulations \cite{White89} remain a tool of choice. Such simulations have
allowed us, for example, to choose between various analytical approaches
that were giving different answers to the pseudogap question in weak  to
intermediate coupling \cite{Moukouri00}.

How can we understand electronic systems that show both localized and
propagating character? Why do both organic and high-temperature
superconductors show broken-symmetry states where mean-field-like
quasiparticles seem to reappear? Why is the condensate fraction in this
case smaller than what would be expected from the shape of the would-be
Fermi surface in the normal state? Are there new elementary excitations that
could summarize and explain in a simple way the anomalous properties of
these systems? Do quantum critical points play an important role in the
Physics of these systems? Are there new types of broken symmetries?
How do  we build a theoretical approach that can include both
strong-coupling and $d=2$ fluctuation effects? What is the origin of
$d$-wave superconductivity in the high-temperature superconductors?  These
are but a few of the basic open questions left to answer in this field.

This work was supported by grants from the Natural Sciences and Engineering
Research Council (NSERC) of Canada and the Fonds pour la formation de
Chercheurs et l'Aide \`a la Recherche (FCAR) of the Qu\'ebec government.
A.-M.S.T. thanks the Issac Newton Institute for Mathematical Sciences,
Cambridge, for hospitality during the writing of this paper.

%============================================================================

%============================================================================


\begin{thebibliography}{10}

\bibitem{Mattheiss}  L.F. Mattheiss, Phys. Rev. Lett. {\bf 58}, 1028 (1987).

\bibitem{Claessen92}  R. Claessen, R.O. Anderson, J.W. Allen, C.G. Olson, C.
Janowitz, W.P. Ellis, S. Harm, M. Kalning, R. Manzke, and M. Skibowski,
Phys. Rev. Lett. {\bf 69}, 808 (1992).

\bibitem{Marshall96} D.S. Marshall, D.S. Dessau, A.G. Loeser, C.-H. Park,
A.Y. Matsuura, J.N. Eckstein, I. Bozovic, P. Fournier, A. Kapitulnik, W.E.
Spicer, and Z.X. Shen, Phys. Rev. Lett. {\bf 76}, 4841 (1996).

\bibitem{Kohn99}  See the Nobel lecture by W. Kohn in Rev. Mod. Phys. {\bf
71}, 1253 (1999).

\bibitem{RevueCBjerome99}  C. Bourbonnais and D. J\'erome, in {\it Advances
in Synthetic Metals, Twenty Years of Progress in Science and Technology,}
edited by P. Bernier, S. Lefrant, and G. Bidan (Elsevier, New York, 1999),
pp. 206-301 (cond-mat/9903101).

\bibitem{Bourbon93} C. Bourbonnais J. Phys. I (France) {\bf 3}, 143 (1993);
P. Wzietek, F. Creuzet, C. Bourbonnais, D.~J\'er\^ome, K. Bechgaard, J.
Phys. I (France) {\bf 3}, 171 (1993).

\bibitem{Bourbon88} C. Bourbonnais, P. Wzietek, D. J\'erome, F. Creuzet, L.
Valade and P. Casssoux, Europhys. Lett., {\bf 6} 177 (1988).

\bibitem{boso}  V. J. Emery in {\it Highly Conducting One-Dimensional
Solids}, eds J. T. Devreese, R. P. Evrard et V. E. van Doren (Plenum 1979),
p. 247. For a recent introduction to bosonization, see D.
S\'en\'echal, cond-mat/9908262 (unpublished).

\bibitem{Haldane81}  F.D.M. Haldane, J. Phys. C{\bf 14}, 2585 (1981).

\bibitem{Lemay}  F. Lemay, unpublished

\bibitem{Vilk97}  Y. M. Vilk and A.-M. S. Tremblay, J. Phys. I France {\bf
7}, 1309 (1997).

\bibitem{Dispute} There is still some dispute in the litterature
on the existence or not of a first-order Mott transition in $d=\infty$
models, but the balance seems in favor of a positive answer to this question.


\bibitem{DMF}  Antoine Georges, Gabriel Kotliar, Werner Krauth and Marcelo
J. Rozenberg, Rev. Mod. Phys. {\bf 68}, 13 (1996).

\bibitem{Pairault}  St\'ephane Pairault, David S\'en\'echal, and A.-M. S.
Tremblay, Phys. Rev. Lett. {\bf 80}, 5389 (1998); also Eur. Phys. J. B, to
appear, cond-mat/9905242.

\bibitem{Long} M. Long, Newton Institute preprint, unpublished.

\bibitem{Lefebvre00} S.~Lefebvre, P.~Wzietek, S. Brown, C. Bourbonnais,
D.~J\'er\^ome,  C. M\'ezi\`ere, M.~Fourmigu\'e and P. Batail, cond-mat/0004455.

\bibitem{Moukouri00}  S. Moukouri, S. Allen, F. Lemay, B. Kyung, D. Poulin,
Y.M. Vilk and A.-M. S. Tremblay, Phys. Rev. B {\bf 61}, 7887 (2000).

\bibitem{Ha96}  For a review, see Z.N.C. Ha, {\it Quantum Many-Body Systems
in One Dimension}, World Scientific, 1996.

\bibitem{BourbonnaisCaron}  C. Bourbonnais and L.G. Caron, Int. J. Mod.
Phys. B {\bf 5}, 1033 (1991);  C. Bourbonnais, in {\it Strongly Interacting
Fermions and High-$T_{c}$ Superconductivity, }Les Houches, Session LVI,
1991, Eds. B. Dou\c{c}ot and J. Zinn-Justin (Elsevier, Amsterdam, 1995),
p.307

\bibitem{CFT}  H. Frahm and V. E. Korepin, Phys. Rev. B{\bf 42}, 10553
(1990); H. Frahm and V. E. Korepin, Phys. Rev. B {\bf 43}, 5653 (1991). For
an introduction to conformal methods, see {\it Conformal Field Theory}, by
P. DiFrancesco, P. Mathieu and D. S\'en\'echal, (Springer Verlag, New York,
1997).

\bibitem{SlaveBoson}  For reviews of slave-boson approaches applied to the
Hubbard model, see for example, G. Kotliar in {\it Correlated Electrons
Systems,} Jerusalem Winter School for Theoretical Physics, Vol.9, Ed. V.J.
Emery (World Scientific, Singapore, 1993), p.118, and G. Kotliar in {\it
Strongly Interacting Fermions and High $T_{c}$ Superconductivity,} Les
Houches, Session LVI, 1991, Eds. B. Dou\c{c}ot and J. Zinn-Justin (Elsevier,
Amsterdam, 1995), p.197.

\bibitem{bickers1}  N. E. Bickers and D. J. Scalapino, Annals of Physics,
{\bf 193}, 206 (1989); N. E. Bickers, D. J. Scalapino and S. R. White, Phys.
Rev. Lett. {\bf 62}, 961 (1989).

\bibitem{Senechal00} Exact diagonalizations can however be combined with
strong-coupling perturbation theory and this attenuates finite-size effects.
See D. S\'en\'echal, D. Perez and M. Pioro-Ladri\`ere, Phys. Rev. Lett. {\bf
84}, 522 (2000).

\bibitem{DMRG}  S.R. White, Phys. Rev. Lett. {\bf 69}, 2863 (1992); Phys.
Rev. B {\bf 48}, 10345 (1993). See also I. Peschel et al.,
{\it Density matrix renormalization}, Lecture notes in physics, vol. {\bf 528},
New York, Springer Verlag (1999).

\bibitem{White89}  S. R. White, D.J. Scalapino, and R. L. Sugar, E. Y. Loh,
Jr., J. E. Gubernatis, and R. T. Scalettar, Phys. Rev. B{\bf 40}, 506 (1989).

\end{thebibliography}
\end{document}